\documentclass[graybox]{svmult}

\usepackage{mathptmx}       
\usepackage{helvet}         
\usepackage{courier}        
\usepackage{type1cm}        
\usepackage{makeidx}         
\usepackage{graphicx}        
\usepackage{multicol}        
\usepackage[bottom]{footmisc}

\makeindex             

\begin{document}

\title*{A new approach to the optimization of the extraction of astrometric and photometric information from multi-wavelength images in  cosmological fields.}
\author{}

\institute{Maria Jose Marquez, \email{mariajose.marquez@eumetsat.int}}
%
%
\maketitle

\abstract{This paper describes a new approach to the optimization of
  information extraction in multi-wavelength image cubes of
  cosmological fields.\newline The objective is to create a framework
  for the automatic identification and tagging of sources according to
  various criteria (isolated source, partially overlapped, fully
  overlapped, cross-matched, etc) and to set the basis for the
  automatic production of the SEDs (spectral energy distributions) for
  all objects detected in the many multi-wavelength images in
  cosmological fields.\newline In order to do so, a processing
  pipeline is designed that combines Voronoi tessellation, Bayesian
  cross-matching, and active contours to create a graph-based
  representation of the cross-match probabilities. This pipeline
  produces a set of SEDs with quality tags suitable for the
  application of already-proven data mining methods.\newline The
  pipeline briefly described here is also applicable to other
  astrophysical scenarios such as star forming regions.}

\section{Introduction}
\label{sec:1}

Single-field multi-wavelength studies obtained with very heterogeneous
instruments and telescopes are very common nowadays. Deep cosmological
surveys are extreme examples of such studies that combine photometric
data from the $\gamma$-rays to the radio-wavelengths, offering
complementary yet astonishingly different views of the same
extragalactic objects. These image cubes carry both astrometric and
photometric information of tens of thousands of sources, which bring
their analysis into the realm of statistics and data mining.

One of the key aspects of the systematic analysis of these image cubes
is the reliability of the scientific products derived from them. In
this work, we concentrate on the generation of spectral energy
distributions (SEDs) of extragalactic sources in deep cosmological
fields. The techniques outlined here are nevertheless of much wider
application in other astrophysical scenarios. We concentrate in
particular in the problem of tagging the quality of a derived SED from
the perspective of the underlying cross-match decisions.

In Section \ref{sec:2} we describe the project and its aims, and the techniques
utilized to derive spectral energy distributions from deep
cosmological image cubes; Section \ref{sec:3} briefly summarizes
the Bayesian approach that serves as the basis for the developments
presented in Section \ref{sec:4} which introduces the possibility of
non-detections in the Bayesian formalism. Finally, Section
\ref{sec:5} describes the results obtained for the application of
the extended formalism to a toy problem, and Section 6 summarizes the
main conclusions.

\section{Deep cosmological fields: the analysis pipeline}
\label{sec:2}
In this work we address the problem of deriving spectral energy
distributions and the labelling of the different sources detected in
multi-wavelength deep images of cosmological fields. Is is compounded
of several sub-tasks, such as the cross-matching of the sources
detected in individual images, the tagging of potential overlaps and
the derivation of optimal regions for sky subtraction. 

Images of the same field obtained with different spatial resolutions,
sensitivities and in various wavelengths will offer complementary
views of the same sources, but also views that can be inconsistent if
we do not take into account all these factors. Let us take for example the
case where a galaxy A detected in low resolution infrared bands has a
flux density below the detection threshold of a mid-infrarred survey,
and has several potential counterparts in visible wavelengths, many of
which do not actually correspond to galaxy A, but to galaxies close to
the line of sight. In addition to this, let us consider the possibility where
one of the visible counterparts (but not the source that corresponds
to galaxy A) is actually detected in the mid-infrared image. A sound
cross-matching approach must necessarily address this problem in a
probabilistic manner, including a requirement on astrometric and
photometric consistency. The approach that we propose here is based on
a Bayesian formalism of  the problem of cross-matching catalogues that,
as a by-product produces a quantitative measure of the validity of the
counterpart assignment and flags SEDs that may be affected by source
overlapping within and across images taken in several bands.

In the first stage of our analysis pipeline, the catalogue extraction
tool Sextractor \cite{Astron.Astrophys.Suppl.Ser:117.393-404} is applied to each image separately. The
catalogue thus obtained (including astrometric and photometric
information) is used as the basis for a 2D Voronoi (Delaunay)
tessellation of the images that defines a polygon in the corresponding
coordinates (e.g., celestial, pixel) for each source.

This 2D Voronoi tessellation of the images provides us with a
preliminary categorization of sources into the candidate categories of
isolated source and partially or totally contaminated by neighbouring
sources. A source is labelled as candidate for isolated source if it
is fully contained in its Voronoi cell and none of the sources from the Voronoi cells surrounding the source under
consideration is contaminating it.  In this initial stage, the source extension is defined
by its Kron ellipse \cite{Astron.Astrophys.Suppl.Ser:117.393-404} although subsequent refinements can be
applied with more refined contours (active contours for example). This
labelling procedure only considers information from one single
image. The definition can be extended by defining an isolated source
as one which is i) isolated in the lowest resolution image; ii) only
has one counterpart in the projection of its Voronoi cell in all other
images and, iii) each of these counterpars is also isolated in the
sense defined above.

\begin{figure}[b]
\label{fig1}
  \begin{center}
 \includegraphics[scale=.55]{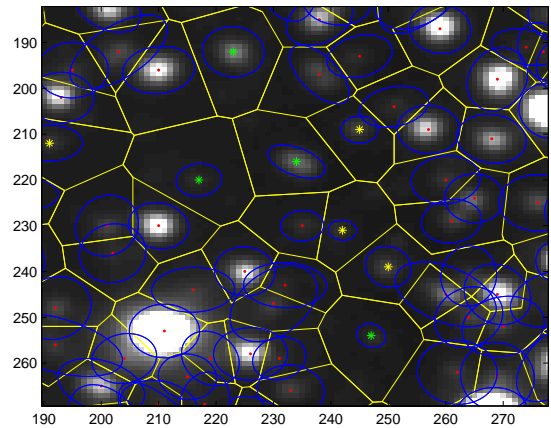}\caption{ Examples of isolated
   and partially contaminated sources in the Hubble Deep Field image
   of IRAC instrument on channel $3.6\mu m$. \label{tesselation}}
\end{center}
\end{figure}

Figure \ref{tesselation} shows an example of the result of the
implementation of this preliminary labelling process to the Hubble
Deep Field image taken by the IRAC instrument on channel $3.6\mu m$.

A simple improvement of this approach consists in taking into account
the source morphology in the determination of the isolation cell by
applying Support Vector Machines for the determination of the maximum
margin hyperplanes separating sources.


The result from the previous steps will produce a set of
two-dimensional vectors, $\vec{x_{ij}}$ ,which represent the celestial
coordinates of the source $j$ in catalogue $i$ together with the
preliminary labelling described in the previous paragraphs. From this
set of vectors, we aim at constructing reliable SEDs by cross-matching
them taking into account the astrometric information, the photometric
information and the instrument sensitivities. In the following, we
will summarize the Bayesian formalism developed in \cite{APJ.679:301-309} that we
further extend to potential non-detections.

\section{Cross-Matching of multi-wavelength astronomical sources}
\label{sec:3}

The work presented in \cite{APJ.679:301-309}, and summarized in the following paragraphs,  proposes a bayesian approach for the decision-making problem of defining counterparts in multi-band
image cubes.

Let us define $M$ as the hypothesis that the position of a source is
on the celestial sphere, and let us parametrize this position in terms
of a three-dimensional normal vector $\vec{m}$. Let us assume that we
have $n$ overlapping images of a given field, and let us call data
$D=\{\vec{x_{1}},\vec{x_{2}},...,\vec{x_{n}}\}$ the $n$-tuple composed
of the locations of $n$ sources in the sky from the $n$ different
channels or images.  Then, two hypothesis can be identified in this
context:

\begin{itemize}
 \item 
$H$: hypothesis that the positions in the $n$-tuple correspond to a
   single source.
 \item 
$K$: hypothesis that the positions do not correspond to a single source.
\end{itemize}

Hypothesis H will be parametrized by a single common location
$\vec{m}$ and the alternative hypothesis K will be parametrized by $n$
positions $\{\vec{m_i}, i:1,2,...,n\}$.

Therefore:
\begin{equation}
\label{eq2}
 P(D|H) =  \int\int\int p(\vec{m}|H) \cdot(\prod_{i=1}^{n}p(\vec{x_{i}}|\vec{m},H)) \,d^3m
\end{equation}

\begin{equation}
\label{eq3}
 P(D|K) =  \int\int\int (\prod_{i=1}^{n}p(\vec{m_i}|K) \cdotp(\vec{x_{i}}|\vec{m_i},K)) \,d^3m_i
\end{equation}

\begin{figure}[b]
\label{figbayes}
  \begin{center}
 \includegraphics[scale=.55]{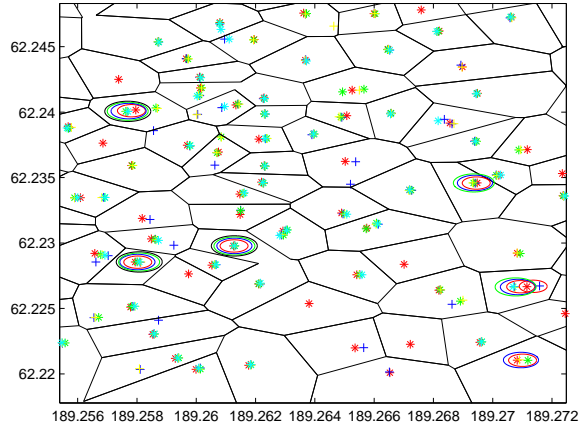}\caption{ Examples of implementation of an iterative procedure for multi-wavelength cross-matching in the Hubble Deep Field image.\label{Bayes circles}}

\end{center}
\end{figure}

In \cite{APJ.679:301-309}, Budavari et al propose an iterative procedure based on the
thresholding of the Bayes factor computed from equations \ref{eq2} and
\ref{eq3} for the identification of counterparts in several
catalogues. We have implemented this procedure and tested it with five
real catalogues (one catalogue from IRAC and four catalogues from
SUBARU). Figure \ref{Bayes circles} shows one example of this implementation. A
threshold of $B_0=5$ was chosen to collect all possible candidates and
the low-probability ones have been weeded out in subsequent steps. A
unique astrometric precision of $\sigma\leq0.2^{''}$ for all
catalogues has been considered.

\section{Extended Bayesian inference for the consideration of non-detection}
\label{sec:4}

The possibility of having non-detected sources has not been taken into
account so far in the formalism described above. In \cite{APJ.679:301-309}, Budavari
et al. suggest one step further by thresholding a combined Bayes
factor that includes the astrometric and the photometric Bayes factors.
In their proposal, the photometric Bayes factor gauges the two hypothesis that i) the photometric
measurements of an $n$-tuple correspond to a single
model SED (where a choice of parameterized models is available for
galactic SEDs), or they come from independent and different SEDs. This
allows us in general to reject a cross-matched proposal, but does not
help in refining it by excluding inconsistent measurements. Here, we
elaborate on that proposal in order to extract that kind of
information that may allow us to construct a SED even if incomplete.

Let us take as starting point $n+1$- tuples derived from the algorithm
proposed in \cite{APJ.679:301-309} which uses only
astrometric information. For obvious reasons, we define the $n+1$-tuple
as a set of potential counterparts to the source detected in the
lowest resolution image which drives the Voronoi tessellation in
celestial coordinates described in section \ref{sec:3}. Let
us define this image as $i=n+1$ in the following. 

In order to include the photometric information into the inference
process, we will assume that there exists a model for the galactic SED
which is parametrized by the set $\{\eta_k, k=1,2,...,K\}$. In \cite{APJ.679:301-309},
  the authors parametrize each SED by a discrete spectral type $T$,
  the redshift $z$ and an overall scaling factor for the brightness,
  $\alpha$; an additional simplification which makes $\alpha=1$ can be
  obtained here by normalizing the SED.

It is important to note that each instrument has its own detection
limit which depends, in general and amongst other factors, on the
spatial flux density of a source and not on the total integrated flux; however, and for the sake of simplicity we will only consider here
flux thresholds instead of fully modelling the detection process, which is always the correct approach, specially when dealing with
extended sources.

The cross-matching problem described in section~\ref{sec:3} requires the ability to identify the same source across different images with different measurement instruments.
The consideration of having sources not detected under study has not been taken into account so far for the model described in \cite{APJ.679:301-309}.

Let us take as starting point $N$ tuples of n$+1$ elements derived from the algorithm proposed in \cite{APJ.679:301-309}.

For the sake of simplicity of this preliminary model,  the existence of one and only one detected source in the channel which drives the voronoi tessellation in celestial coordinates described in section \ref{sec:3} will be assumed, therefore there will always exist a detection in this channel.

To deal with the concept of non-detection, the use of photometric information is required and for that purpose the photometric model proposed in \cite{APJ.679:301-309} will be used and extended. As indicated in \cite{arXiv:astro-ph.CO/1008.0395v1}, a wealth of models has been created with the goal of choosing and extracting useful information from SEDs.
In our case we will follow the same simple model for the SED as the one indicated in \cite{APJ.679:301-309}.
Let us consider the data $D'$ as an n$+1$-tuple of the measured fluxes: $D'=\{\vec{g_1},\vec{g_2},...,,\vec{g_{n+1}}\}$.

The Bayesian inference for this photometric model will be run on the following two mutually exclusive hypothesis:
\begin{itemize}
 \item 
H$_1$: all the fluxes $\vec{g_i}$ correspond to the same source.
\item
K$_1$: not all the fluxes $\vec{g_i}$ correspond to the same source.
\end{itemize}

The evidences for the hypothesis H$_1$ and K$_1$ are:

\begin{equation}
\label{eq4}
p(D'|H_1) = \int p(\vec{\eta}|H_1)\prod_{i=1}^{n+1}p_i(\vec{g_i}|\vec{\eta},H_1)d^r\eta
\end{equation}

where:
\begin{itemize}
\item
$\vec{\eta}$ are the parameters for modelling the spectral energy distribution.
\item
$p(\vec{\eta}|H_1)$ is the prior probability which should be carefully chosen from one of the models proposed in \cite{arXiv:astro-ph.CO/1008.0395v1}, for example, SWIRE database could be a good option for IRAC catalogues.
\item
$p_i(\vec{g_i}|\vec{\eta},H_1)$ is the probability that one source with SED parameters $\vec{\eta}$ has a measured flux of $\vec{g_i}$.
\end{itemize}

 For hypothesis K$_1$ we will take on board the consideration for the possibilities of having sources non-detected in one or several channels. This means that  the hypothesis K$_1$ contains a combinatorial number of sub-hypothesis (i.e. that the source has not been detected in any possible combination of channels, and that the detections in these channels correspond to nearby sources in the celestial sphere).

In this way, one new sub-hypothesis is established per combination found; therefore there will be:
\begin{itemize}
 \item 
$C_{n,1}=n$ sub-hypothesis for one non-detection.
\item
$C_{n,p}=\frac{n!}{p!\cdot(n-p)!}$ sub-hypothesis for p non-detections.
\item
$C_{n,n} =1$ sub-hypothesis for n non-detections.
\end{itemize}

The formalism proposed here for the hypotheis K$_1$ will include all the independent sub-hypothesis described before. \newline

Let us be $P_{n,p}=\{L_{\{i_1,...,i_p\}}\}$ the set of sub-hypothesis with $p$ non-detections and with $n-p$ detections.

The generic expression for hypothesis K$_1$, taking on board all the possibilities of non-detections from an n+1-tuple is as follows: 
\begin{eqnarray}
\label{eq5}
p(D'|K_1) = \prod_{i=1}^{n+1}\left\{\int p(\vec{\eta_{i}}|L)\cdot p_i(\vec{g_i}|\vec{\eta_i},L)d^r\eta_i\right\} \\ \nonumber
= \sum_{p=1}^{n}\sum_{ L\in P_{n,p}}\prod_{ i=\{i_1,...,i_p\}}\left\{\int\int_{-\infty}^{\theta_{thi}} p_i(\vec{g_i}|\eta_i,L)p(\vec{\eta_i}|L)d\vec{g_i}d^r\eta_i\right\}\cdot\prod_{j=1,j\neq i}^{j=n+1}\left\{\int p(\vec{\eta}|L)\cdot p_j(\vec{g_j}|\vec{\eta},L)d^r\eta \right\}
\end{eqnarray}

Where the non-detection for the source $i$ can be modelled as the area below the detection threshold, $\theta_{thi}$, of a Gaussian distribution.

The evidence for hypothesis K$_1$, as expressed in equation \ref{eq5}, includes the combinatorial number of the exclusive sub-hypothesis presented before. In this way, an unambiguous description of each specific combination of non-detection(s) among the channels of the n$+1$-tuple is feasible.

The use of the different Bayes Factors per sub-hypothesis will allow the identification of the most favourable model; alternatively other statistics as the Bayesian Model Averaging (BMA) can also provide the assessment on how probable is a model given the data conditionally on a set of models considered, L$_1$,...,L$_p$,...,L$_n$, being L$_p$ the set of sub-hypothesis corresponding to $p$ non-detections.
Initially we would assign the same value for each sub-hypothesis.

\section{Toy example}
\label{sec:5}
Let us model the radiation of a black body using Planck's law. This function depends on the frequency $\nu$.
\begin{equation}
\label{eq14}
I(\nu,T) = \frac{2h\nu^5}{c^3}\cdot\frac{1}{e^\frac{h\nu}{KT}-1}
\end{equation}
Let us consider a set of measurements $g_i$ of the black body intensities $I(\nu,T)$, therefore for a 6-tuple we will have the following data: $D'=\{g_1,....g_6\}$;
for the prior we will use a flat function as a first approximation and we will assume a Gaussian distribution for the uncertainties measurement $p_i(g_i|T)$.
Note that in this case the example has been modelled in such a way that the measurement in channel $3$, $ g_3$,  will not correspond to the cross-matching.

Applying equations \ref{eq4} and \ref{eq5} we obtain the following Bayes factor:
\begin{equation}
 \label{eq18}
B = \frac{p(D|H_1)}{p(D|K_1)} = 1.92\cdot10^{-2} 
\end{equation}

Therefore the model K$_1$ will be clearly more favourable than the model H$_1$. We can go a step further by applying here the extended Bayesian formalism presented, from which the Bayes Factors of all possible sub-hypothesis are obtained, resulting sub-hypothesis $L_{\{3\}}$  the most favourable one, as expected.

\begin{figure}
\label{fig3}
 \centering
 \includegraphics[scale=0.45]{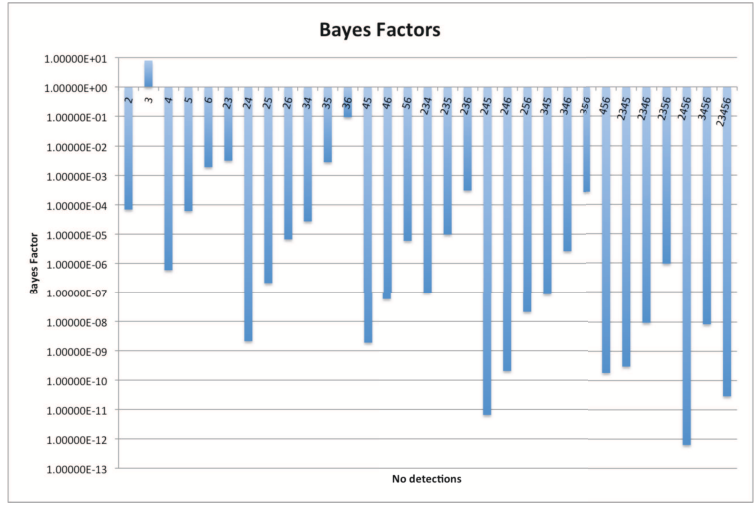}\caption{Bayes Factors for all sub-hypothesis included in the hypothesis $K_1$}
\end{figure}

\section{Conclusions}
\label{sec:6}
The proposed extended Bayesian formalism for the probabilistic cross-matching problem drives the identification of the most favourable model among many when all the posible exclusive combinations of having non-detected sources within the n$+1$-tuple are taken into account; this stage leads to an obvious refinement phase in the construction of consistent SEDs, allowing a more precise labelling process for sources detected in multi-wavelength deep images of cosmological fields.

%
%

%
%
%

\end{document}